\documentclass[12pt,preprint]{aastex}

\newcommand{\Teff}{\mbox{$T_{\rm eff}$}}
\newcommand{\logg}{\mbox{$\log g$}}

\newcommand\vmac{\mbox{$v_{\rm mac}$}}

\newcommand{\vsini}{\mbox{$v\sin i$}}
\newcommand{\kms}{\mbox{km s$^{-1}$}}

\slugcomment{DRAFT}

\begin{document}

\title{\bf Measuring the Magnetic Field on the Classical T Tauri Star \\ TW Hydrae}

\author{Hao Yang}
\affil{Department of Physics \& Astronomy, Rice University, 6100 Main St.
       MS-108, Houston, TX 77005}
\email{haoyang@rice.edu}

\author{Christopher M. Johns-Krull\altaffilmark{1,2}}
\affil{Department of Physics \& Astronomy, Rice University, 6100 Main St.
       MS-108, Houston, TX 77005}
\email{cmj@rice.edu}

\author{Jeff A. Valenti\altaffilmark{1}}
\affil{Space Telescope Science Institute, 3700 San Martin Dr., Baltimore, MD
       21210}
\email{valenti@stsci.edu}

\altaffiltext{1}{Visiting Astronomer, Infrared Telescope Facility, operated
for NASA by the University of Hawaii}
\altaffiltext{2}{Visiting Astronomer, McDonald Observatory, operated
by The University of Texas at Austin}

\begin{abstract}

     We present infrared (IR) and optical echelle spectra of the Classical T Tauri star TW Hydrae. Using the optical data,
we perform detailed spectrum synthesis to fit atomic and molecular absorption lines and determine key stellar parameters:
\Teff\ $=4126 \pm 24$ K, \logg\ $=4.84 \pm 0.16$, [M/H]$=-0.10 \pm 0.12$, \vsini\ $= 5.8\pm 0.6$ \kms.
The IR spectrum is used to look for Zeeman broadening of photospheric absorption lines.  We fit four Zeeman sensitive 
\ion{Ti}{1} lines near $2.2$ microns and find the average value of 
the magnetic field over the entire surface is $\bar B$ = $2.61\pm0.23$ kG. In addition, several nearby magnetically insensitive CO lines show 
no excess broadening above that produced by stellar rotation and instrumental broadening, reinforcing the magnetic 
interpretation for the width of the \ion{Ti}{1} lines. We carry out extensive tests to quantify systematic 
errors in our analysis technique which may result from inaccurate knowledge of the effective temperature or gravity, 
finding that reasonable errors in these quantities produce a $\sim10$\% uncertainty in the mean field measurement. 

\end{abstract}


\section{Introduction}

     T Tauri stars (TTSs) are newly formed low-mass stars that have recently become visible at optical wavelengths. They are 
young, roughly solar mass stars, still contracting along pre-main sequence evolutionary tracks in the H-R diagram. TTSs 
can be placed into two categories, depending on whether or not they are actively accreting. Classical T Tauri stars (CTTSs) 
are still surrounded by disks of material that are undergoing accretion onto the central star, producing
excess emission in the form of both lines and continuum at ultraviolet (UV), optical, infrared (IR), 
and radio wavelengths. Naked T Tauri stars (NTTSs) are those TTS that do not appear to have close dusty accretion disks 
surrounding the central stars and show no indications of accretion onto the star\footnotemark. The properties of TTSs 
have been reviewed by Appenzeller \& Mundt (1989), Bertout (1989), Basri \& Bertout (1993), and M\'enard \& Bertout (1999). 
It is within the disks surrounding CTTSs that planetary systems form. Understanding the processes through which young stars 
interact with and eventually disperse their disks is critical for understanding the formation of planetary systems 
as well as understanding the rotational evolution of young stars.

\footnotetext{The term weak or weak-lined TTSs (WTTSs) is often used to refer to NTTSs; however, WTTS is a designation
based purely on the observed equivalent width of the H$\alpha$ emission line, and there are several examples of WTTS that
are nevertheless accreting material from a circumstellar disk at rates comparable to that found for CTTSs.  Therefore,
we will use NTTS to refer to stars which do not show any sign of accretion.}

     Magnetic fields are believed to play a fundamental role throughout the star formation process, especially 
with regard to the interaction between a CTTS and its circumstellar disk. Stellar 
magnetic fields are thought to truncate the disk near the corotation radius, directing the flow of accreting 
material toward the polar regions of the central star (Camenzind 1990, K\"onigl 1991, Cameron \& Campbell 1993, Shu et al. 
1994, Paatz \& Camenzind 1996). Stellar magnetic fields may also play a critical role in driving the strong
winds/jets seen in young stars. Therefore, a full understanding of the early evolution of young stars requires
exploration of their magnetic field properties.

     The most successful approach for measuring fields on late--type stars in general has been to measure Zeeman broadening 
of spectral lines in unpolarized light (e.g., Robinson 1980; Saar 1988; Valenti, Marcy, \& Basri 1995; Johns--Krull \& 
Valenti 1996).  For any given Zeeman $\sigma$ component, the wavelength shift
caused by a magnetic field is
$$\Delta\lambda = {e \over 4\pi m_ec^2} \lambda^2 g B
                = 4.67 \times 10^{-7} \lambda^2 g B \,\,\,\,\,\,
                {\rm m\AA\ kG}^{-1},\eqno(1)$$
where $g$ is the Land\'e-$g$ factor of the transition, $B$ is the strength of the magnetic field (in kG), and $\lambda$
is the wavelength of the transition (in \AA). The first direct detection of a magnetic field on a TTS was
reported by Basri, Marcy, and Valenti (1992).  These authors used the fact that
moderately saturated magnetically sensitive 
lines will have an increase in their equivalent width in the presence of a magnetic field as the line components split to 
either side, increasing line opacity in the wings.  Using this equivalent width technique, Basri et al. (1992) measured 
the field of the NTTS Tap 35, obtaining a value of the magnetic field strength, $B$, times the filling factor, $f$, of field 
regions of $Bf = 1.0 \pm 0.5$ kG.  As the importance of magnetic fields on TTSs has become recognized, more effort has 
been put into their measurement.  Guenther et al. (1999) used the same equivalent width technique to 
analyze spectra from 5 TTSs, claiming significant field detections for 2 stars.  

     Johns--Krull, Valenti, and Koresko (1999b - hereafter Paper I) have examined three (K\"onigl 1991, Cameron \& Campbell 
1993, Shu et al. 1994) of the magnetospheric accretion theories and provide analytic expressions for 
the equatorial magnetic field strength at the surface of CTTSs.  In the context of these theories, the surface field strength depends on 
the stellar mass, radius, rotation rate, and the mass accretion rate onto the star from the disk. 
In all these theories, a dipole field geometry is assumed.
Paper I shows that the predicted dipole magnetic field strengths vary over a large range of values, 
including strengths up to several kilogauss.
Such high field strengths should be directly measureable as Zeeman broadening in the line profiles of the most magnetically sensitive diagnostics.

     Paper I made use of the wavelength dependence of the Zeeman effect (equation 1) to detect actual Zeeman broadening of 
a \ion{Ti}{1} line at 2.2 $\mu$m on the CTTS BP Tau, measuring a field
strength of $\bar B = 2.6 \pm 0.3$ kG, averaged over the entire surface.  
Johns--Krull, Valenti, and Saar (2004 - hereafter Paper II)  analyzed 4 magnetically sensitive \ion{Ti}{1} lines in
K-band spectra of the NTTS Hubble 4, finding an average field strength of $\bar B = 2.5 \pm 0.2$ kG. Paper II also examined magnetically
insensitive CO lines at 2.3 $\mu$m and detected no excess broadening above that due to rotation and instrumental broadening. 
Papers I and II also point out that the mean fields detected on BP Tau and Hubble 4 are larger than those which can be
confined by pressure balance with the quiet photosphere as discussed by Safier (1999); however, this is not a problem if 
the magnetic filling factor is unity in the visible photosphere.  Indeed, the filling factor is unity in the dipole field 
configurations favored in current theoretical models (K\"onigl 1991, Cameron \& Campbell 1993, Shu et al. 1994).

     Another method for detecting magnetic fields is to look for net circular polarization in Zeeman sensitive lines.  
Observed along a line parallel to the magnetic field, the Zeeman $\sigma$ components are circularly polarized, with 
opposite helicity on either side of the nominal line wavelength. If the vector sum of magnetic fields on the stellar
surface has a significant component along the line of sight, then net circular polarization becomes detectable.
If the magnetic topology on the star is complex 
and small scale (as on the Sun), nearly equal amounts of both field polarities will generally be present producing 
negligible net polarization. Johns--Krull et al. (1999a) and Daou, Johns--Krull, and Valenti (2005) fail to detect 
polarization
in photospheric absorption lines at 3$\sigma$ upper limits of 200 G and 105 G respectively, seriously calling into question
the assumed dipole field geometry. On the other hand, Johns--Krull et al. (1999a) discovered net polarization in the 
\ion{He}{1} $\lambda$5876 {\it emission} line on the CTTS BP Tau, indicating a net longitudinal magnetic field of 
$2.46 \pm 0.12$ kG in the line formation region.  This \ion{He}{1} emission line is believed to form in the post-shock 
region where disk material accretes onto the stellar surface (Hartmann, Hewett, \& Calvet 1994; Edwards et al. 1994). 
Circular polarization in the \ion{He}{1} line has now been observed in several CTTSs (Valenti \& Johns--Krull 2004;
Symington et al. 2005).  These observations suggest that accretion onto CTTSs is indeed controlled by a strong stellar 
magnetic field.
  
While it is clear that at least a few TTSs possess kilogauss fields over most or all of their surfaces, it is also
quite likely these fields are not dipolar.  More observation of photospheric magnetic fields on TTSs are needed to
test the assumptions and predictions of magnetospheric accretion theories.  Here, we present an analysis of the
photospheric magnetic field on the CTTS TW Hya, which is a K7Ve(Li) star (Herbig 1978) located at a distance of
$56 \pm 7$ pc (Wichmann et al. 1998) in the TW Hydrae association.  TW Hya is 13$^\circ$ from the nearest dark cloud,
but it nonetheless has a disk with a mass of $\sim 5\times 10^{-3} {\rm M}_{\odot}$, as inferred from dust emission
(Wilner et al. 2003).  According to images at visible, near-IR, millimeter and sub-millimeter wavelengths (Krist et
al. 2000; Wilner et al. 2000; Alencar \& Batalha 2002; Weinberger et al. 2002; Qi et al. 2004), the optically thick
disk is seen nearly face-on, extending to a radius of at least  $3\farcs5$ (200 AU).  Rucinski \& Krautter (1983)
find that TW Hya has all the properties of a classical T Tauri star, such as strong and variable Balmer emission
lines (see also Webb et al. 1999) and a strong IR excess (see also Jayawardhana et al. 1999).  Strong UV excess
emission suggests ongoing accretion onto the star with an accretion rate estimated at $2 \times 10^{-9}
{\rm M}_\odot {\rm yr}^{-1}$ (Herczeg et al. 2004).  Age estimates for TW Hya range from $7 - 15$ Myrs (Hoff et
al. 1998, Webb et al. 1999, van Dishoeck et al. 2003).

Below, we report detection of Zeeman broadening in four \ion{Ti}{1} lines in the K band, indicating that a strong
magnetic field covers most or all of the surface of TW Hya. To further verify the magnetic nature of the detected 
broadening, several magnetically insensitive CO lines are also observed in the IR, exhibiting no broadening beyond that 
expected from stellar rotation as determined from an analysis of the optical spectrum. 
We also present an extensive analysis of possible systematic errors, finding that 
reasonable errors in effective temperature and gravity contribute only a $\sim$10\% uncertainty 
to the derived mean field. The observations and data reduction are described in \S\ 2. In \S\ 3 the analysis
of the data is presented: \S\ 3.1 examines the optical data used to determine the basic atmospheric parameters of TW Hya; 
while \S\ 3.2 presents the analysis of the IR data to measure the magnetic properties of this star; and \S\ 3.3 examines 
possible systematic errors in models used to analyze the IR data. A discussion of the results is given in \S\ 4, and 
our conclusions are summarized in \S\ 5.

\section{Observations and Data Reduction}

     We obtained optical spectra at the 2.1 m Otto Struve telescope at McDonald Observatory on 1999 February 5 and 6. 
A Reticon $1200 \times 400$ CCD was used to record each spectrum covering the wavelength interval 5803 -- 7376 \AA. 
The spectrograph slit was approximately 1\farcs07 arcsec wide on the sky, which projected to $\sim 2.14$ pixels on the detector, 
resulting in a spectral resolving power of $R \equiv \lambda/\delta\lambda = 56,000$. These spectra were reduced using 
an automated echelle reduction package written in IDL that is described by Hinkle et al. (2000). Data reduction included 
bias subtraction, flat-fielding by a normalized incandescent lamp spectrum, scattered light subtraction, cosmic ray removal, and optimal
extraction of the spectrum. Wavelengths were determined by fitting a two-dimensional polynomial to $n\lambda$ as function 
of pixel and order number, $n$, for several hundred thorium lines observed in an internal (part of the spectrograph assembly
-- post telescope) lamp. In addition to observations of TW Hya, spectra were also obtained of a rapidly rotating hot 
star, $\alpha$ Leo, at similar airmass to correct for contamination by telluric absorption lines. Table \ref{obs} summarizes 
the optical observations.

     The K-band spectra presented here were obtained at the NASA Infrared Telescope Facility (IRTF) in April 1998.  
Observations were made with the CSHELL spectrometer (Tokunaga et al. 1990, Greene et al.\ 1993) using a 0\farcs5 slit to 
achieve a spectral resolving power of $R \sim 36,200$, corresponding to a FWHM of $\sim 2.7$ 
pixels at the $256 \times 256$ InSb array detector.  A continuously variable filter (CVF) isolated individual orders 
of the echelle grating, and a 0.0057 \micron\ portion of the spectrum was recorded in each of 3 settings.  The first two 
settings (2.2218 and 2.2291 $\mu$m) each contain 2 magnetically sensitive \ion{Ti}{1} lines.  The third setting (2.3125 
$\mu$m) contains several strong, magnetically insensitive CO lines. Each star was observed at two positions along the slit 
separated by 10\arcsec.  The K band spectra were reduced using custom IDL software described in Paper I. The reduction 
includes removal of the detector bias, dark current, and night sky emission. Flat-fielding, cosmic ray removal, and optimal
extraction of the stellar spectrum are also included. Wavelength calibration for each setting is based on a 3rd order 
polynomial fitted to $n\lambda$ for several lamp emission lines observed by changing the CVF while keeping the grating 
position fixed. Dividing by the order number for each setting yields the actual wavelength scale for the setting in question.
In each wavelength setting, we observed TW Hya and $\alpha$ Leo. The IR observations are also summarized in Table \ref{obs}.

\section{Analysis}

     The \ion{Ti}{1} lines in the K band are excellent probes of magnetic fields on cool stars because of the $\lambda^2$ 
wavelength dependence of the Zeeman effect (equation 1) compared to the $\lambda$ dependence of Doppler broadening.
Here we measure the magnetic field on TW Hya by modeling the profiles of K band absorption lines. Since other line 
broadening mechanisms such as rotation are also potentially important, we must first demonstrate that we can accurately 
predict the appearance of the spectrum in the absence of a magnetic field. Several steps are required to achieve this goal, 
and the work here builds on the results of Papers I and II.

     First, the optical spectrum of TW Hya is fit assuming no Zeeman broadening, since rotational and turbulent broadening 
dominate at these wavelengths. Based on fits to the solar spectrum in Paper I, we have precise atomic line data for the 
regions of the spectrum used here to determine atmospheric parameters. Fitting the optical spectrum of TW Hya yields the 
stellar parameters \Teff, \logg, [M/H], \vsini\, and veiling (\textit{r}). Using the resulting parameters, the nonmagnetic 
IR spectrum is predicted, and comparison with the observations shows obvious excess broadening in the observed 
\ion{Ti}{1} line profiles, while no such excess broadening is seen in the magnetically insensitive CO lines. This excess 
width in the \ion{Ti}{1} lines is then modeled as Zeeman broadening, using a polarized radiative transfer code. Matching 
the observed \ion{Ti}{1} profiles then gives the strength and filling factor of magnetic regions on the surface of TW Hya.

\subsection{Fit to the Optical Spectrum}

     The details of the spectrum synthesis and fitting procedure are given in Paper I. Briefly, spectra for the optical 
wavelength regions are calculated on a grid (in \Teff, \logg, and [M/H]) of model atmospheres. The nonlinear least 
squares technique of Marquardt (see Bevington \& Robinson 1992) is used to fit the observed spectrum, solving for the key
stellar parameters, namely, 
\Teff, \logg, [M/H], \vsini, and \textit{r}. Spectra between model grid points are computed using three-dimensional 
linear (with respect to the model parameters\ \Teff, \logg \ and [M/H]) interpolation of the resulting spectra. The model 
atmospheres are the ``next generation'' (NextGen) atmospheres of Allard \& Hauschildt (1995). We interpolate the logarithm 
of the line depth rather than residual intensity, because the former quantity varies more slowly with changes in model 
parameters, yielding more accurate interpolated values. Rotational broadening is larger than the effects of macroturbulence 
in TW Hya, which makes it difficult to solve for \vmac. We adopt a fixed \vmac\ of 2 \kms\ from an extrapolation to K7 of 
class IV stars given in Gray (1992). It was found by Valenti et al.\ (1998) that in M dwarfs, microturbulence and 
macroturbulence were degenerate for the low trubulent velocities considered here. Therefore, microturbulence is neglected, 
allowing \vmac\ to be a proxy for all turbulent broadening in the photosphere. Spectra are synthesized in 3 optical regions,
centered on 6151 \AA, 6498 \AA, and on the TiO bandhead at 7055 \AA, which provides excellent temperature sensitivity. 
Following Paper I, we only use features which are modeled well in both the solar spectrum and the spectrum of 61 Cyg B, 
a K7V reference star.

\begin{figure}[ht]
  \begin{center}
  \includegraphics[scale=0.65]{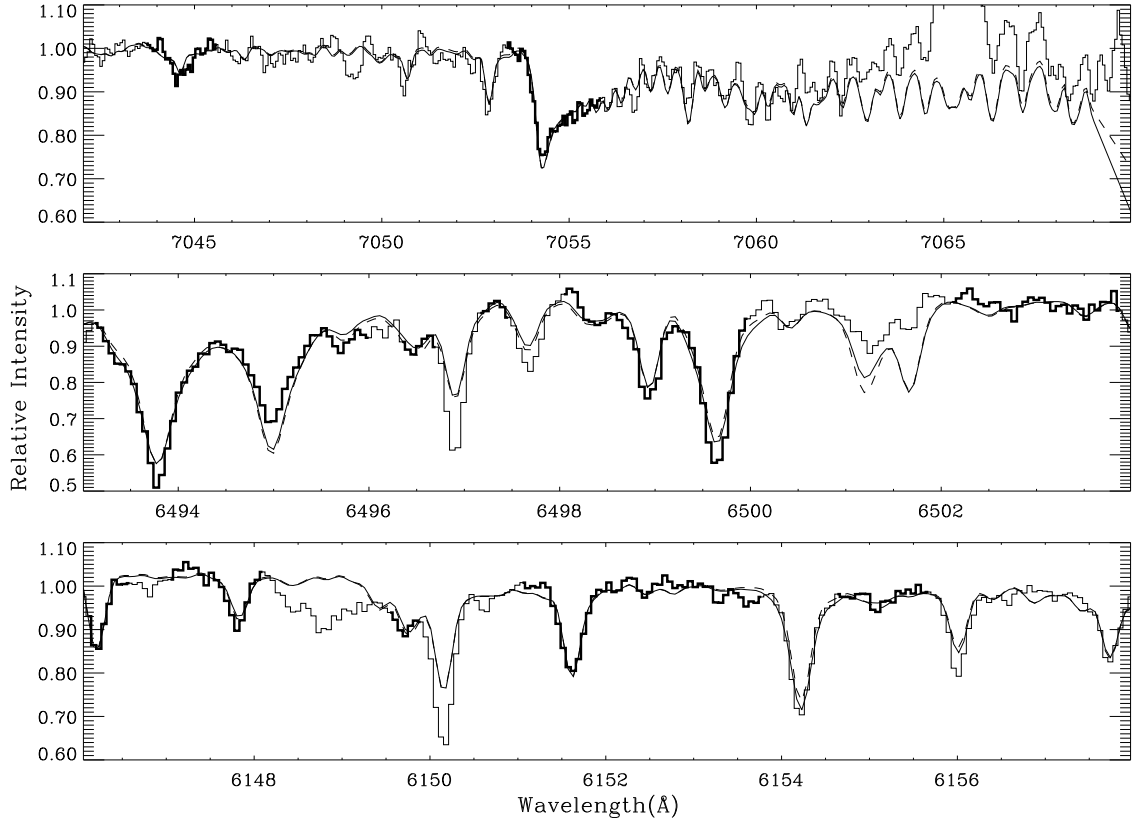}
  \end{center}
\caption{Best fit to the optical spectra
     (histogram : data ; smooth solid line : best fit with \logg\ as a free parameter ; 
     dash-dotted line : best fit with \logg\ fixed at 4.38). 
     The thick regions of the histogram are the spectral regions
     actually used in the fit.} \label{opticfig}
\end{figure}

TW Hya model parameters determined by fitting the optical spectrum are given in the first row of Table \ref{optictab}. 
Observed and synthetic spectra for the three wavelength regions are shown in Figure \ref{opticfig}. The Marquardt technique 
used to find the best fitting stellar parameters also gives formal uncertainties, but the actual uncertainties in derived
parameters are completely dominated by systematic errors resulting from residual uncertainties in the line data 
and the model atmospheres used in the fit.  To estimate these uncertainties we reanalyzed the observed spectra using 9
out of 10 atomic line intervals between 6140 - 6500\AA\ (indicated by contiguous bold segments in the bottom two panels
of Figure 1), cycling through all 10 combinations.
This procedure yields 10 alternate values for each model parameter and the standard deviation of each parameter value 
provides a more realistic estimate of its uncertainty. 
These uncertainties  are listed in the second row of Table \ref{optictab}. 
The order containing the 7055 \AA\ molecular band head is always included in this procedure 
because this band head is such a strong temperature constraint.  It is likely that our estimate of the systematic
uncertainty on the \Teff\ determination is an underestimate.

\subsection{Fits to the Infrared Data}

     Using the parameters (\Teff, \logg, [M/H], and \vsini) determined by the optical fits, we fit the 
magnetically insensitive CO lines at 2.3 $\mu$m and the magnetically sensitive \ion{Ti}{1} lines at 2.2 $\mu$m. 
The CO lines serve as a check that the parameters determined in the optical accurately represent the IR spectrum
(particularly the line width), while the \ion{Ti}{1} lines 
allow a determination of the magnetic field on TW Hya. The two categories of IR data are discussed separately below.
Johns--Krull and Valenti (2001) find no K band veiling for this spectrum of TW Hya. Therefore, no veiling is used in the fits 
to the IR data for both the CO and \ion{Ti}{1} line regions.

\subsubsection{Magnetically Insensitive CO Lines}

\begin{figure}[ht]
 \begin{center}
  \includegraphics[scale=0.60]{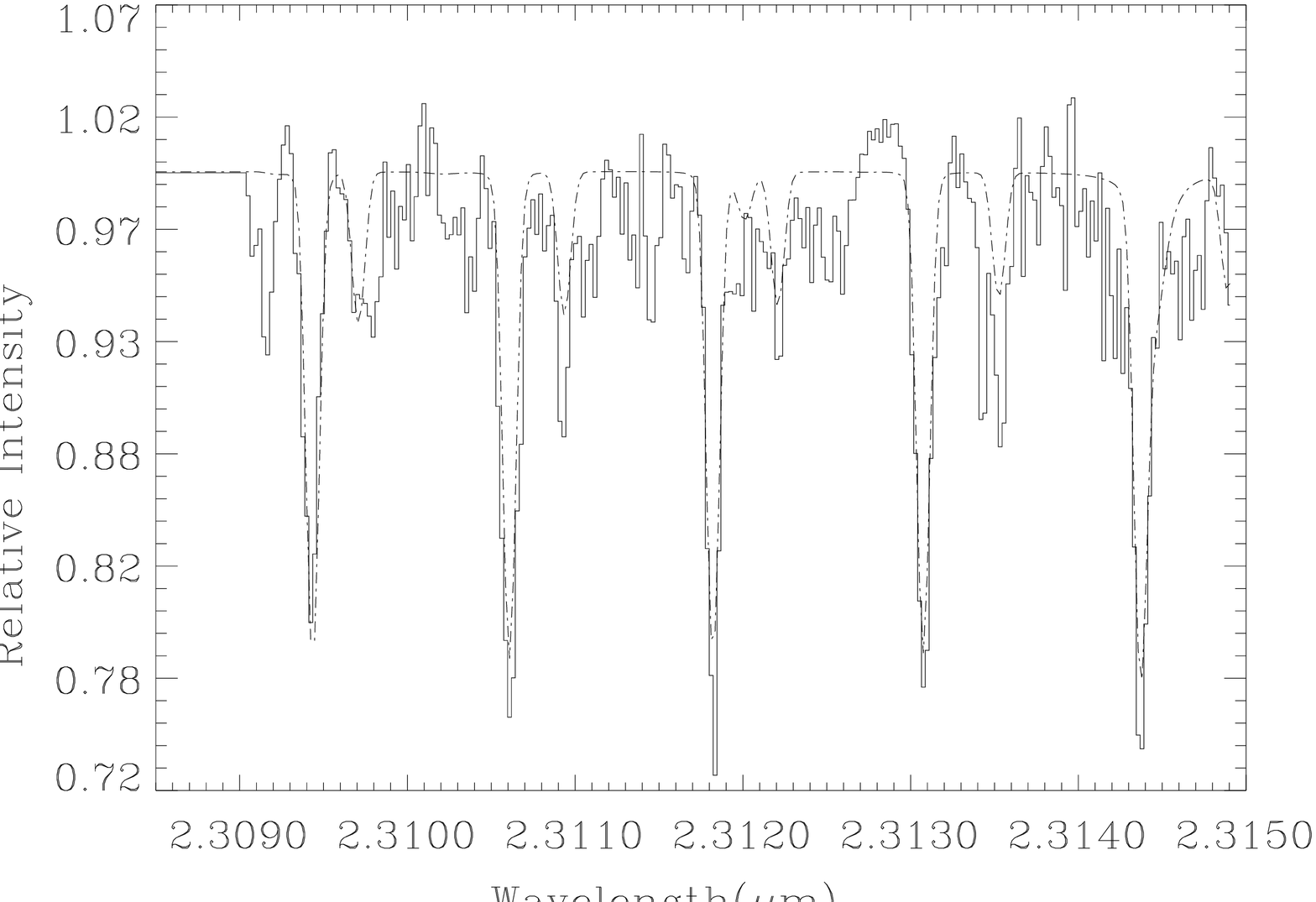}
  \end{center}
  \caption {The solid histogram shows the observed CO lines of TW Hya. The dash-dotted
   curve shows the model spectrum produced by fitting the optical spectrum. No fitting was
   done to the CO lines.} \label{cofig}
\end{figure}

     Line data for CO is taken from Goorvitch (1994) for the CO {\it X}$\Sigma^+$ state. We constructed a model atmosphere
by interpolating a grid of NextGen model atmospheres to the \Teff, \logg, and [M/H] obtained in the optical analysis.  We
then synthesized the CO line spectrum and convolved with a Gaussian corresponding to the spectral resolving power of the
observed spectrum. Continuum normalization was the only free parameter used to match the observed and synthetic spectra,
which are compared in Figure \ref{cofig}. The shape of the synthetic spectrum is determined entirely by parameters obtained
in the optical analysis. Within the signal-to-noise ratio of the data, the model spectrum in Figure \ref{cofig} shows good
agreement with the observed spectrum, yielding a reduced $\chi^2$ of $\chi^2_r = 2.00$. Line strengths are off slightly,
but this could in principle by corrected by tuning CO oscillator strengths.  However, our primary concern is the widths of
the CO lines, which are accurately reproduced by the synthetic spectrum, demonstrating that we are correctly modeling the
dominant nonmagnetic broadening mechanisms.

\subsubsection{Magnetically Sensitive \ion{Ti}{1} Lines}

\subsubsubsection{Single Field Component Model}

     We start by applying a simple model in which some fraction of the surface is covered by a field of a single strength, 
while the rest of the surface is field free. Such a model is then specified by a single field strength, $B$, and the filling 
factor, $f$, of that field on the surface. By analogy to the mean field direction in solar plage, we assume a radial field 
geometry in the stellar photosphere. We assume that thermal structures are identical in magnetic and field-free (quiet)
regions. This is appropriate for optically thin flux tubes (e.g., plage), but probably not for large spots. The thermal
structure of magnetic regions on TTS is unknown, nor are any applicable models available for use in our analysis.

\begin{figure}[ht]
 \begin{center}
  \includegraphics[scale=0.65]{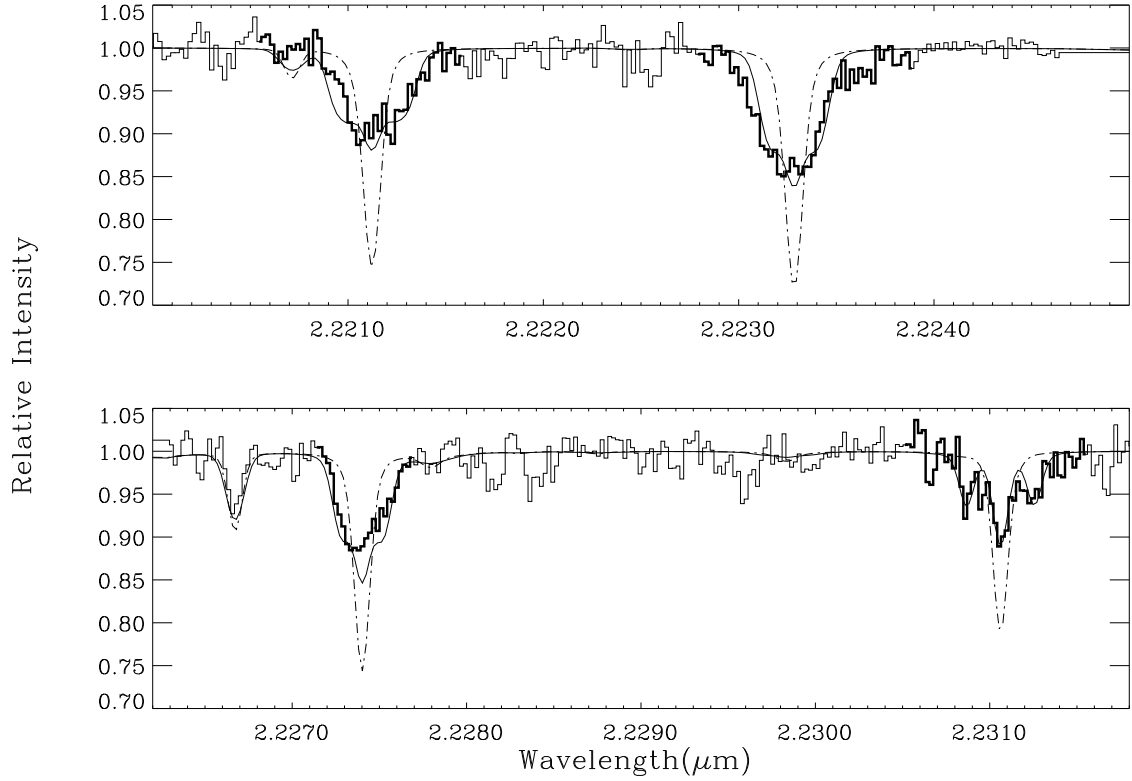}

\caption{Best fit of infrared spectra using a single-component
magnetic field model. Magnetically sensitive Ti I lines are
fitted, where the bold regions indicate the portions of the spectrum
actually used in the fit (histogram: data;
 dash-dotted line: fit without magnetic field;
 smooth solid line: fit with magnetic broadening).
}
  \label{irfig}
  \end{center}
\end{figure}

     The spectrum synthesis including magnetic fields has been described in Paper I. Only one \ion{Ti}{1} line was used to 
model magnetic fields on BP Tau in Paper I. Strong K band veiling in BP Tau masked any line strength 
uncertainty due to errors in atomic data. In the case of 
Hubble 4 in Paper II, there is no K band veiling and all the \ion{Ti}{1} lines used here were available to model. After
comparing with the solar spectrum and that of 61 Cyg B, the $gf$ values of the 4 \ion{Ti}{1} lines were adjusted to match 
the profiles observed in 61 Cyg B. Here for TW Hya, we adopt the $gf$ values for the \ion{Ti}{1} lines from Paper II. 
The Marquardt technique is used again to solve for the best fitting values of the magnetic field strength and the 
filling factor. The key atmospheric parameters (\Teff, \logg, [M/H], and \vsini) are held fixed at values determined from 
the analysis of the optical data. The best fit to the IR data with a single field component model yields $B = 3.35$ kG and 
$f = 0.66$, and this model is shown in Figure \ref{irfig} along with the model assuming no magnetic field on TW Hya. The
model with a magnetic field is clearly a much better fit to the observed
spectrum, yielding a reduced $\chi^2$ of $\chi^2_r = 1.70$ (computed considering only the points in the \ion{Ti}{1} lines
shown in bold in Figure \ref{irfig}). The non-magnetic model with all parameters fixed to values from the optical 
analysis gives $\chi^2_r = 5.63$. For comparison with the models below, we note that $\bar B \equiv Bf = 2.21$ kG for
this model.

\subsubsubsection{Multi-Component Field Models}

     The single field component model provides a decent fit to the observed lines; however, the strengths of 
the lines at 2.22112 and 2.22740 $\mu$m are over-predicted. 
In addtion, the observed profiles are broad and smooth, while the computed profiles show obvious inflection points. 
In Paper I this motivated the authors to construct  multi-component field 
models which provided the best fit to BP Tau \ion{Ti}{1} profile. In Paper II several multi-component field models are
applied and improved the fits as well. It was also found in Paper I that multicomponent models with too many field
components are degenerate since it is difficult to constrain separate field components with relatively small differences in 
field strength due to the combined effects of the spectrometer resolution and stellar rotation. In the case of BP Tau in 
Paper I, the effective magnetic resolution is about 2 kG. For Hubble 4 in Paper II, the high \vsini\ of 14 \kms\ lowers the 
resolution of Zeeman components, but simultaneous fitting of four \ion{Ti}{1} lines with different Land\'e-$g$ values improves the situation, 
resulting in approximately the same
effective magnetic resolution as the analysis of BP Tau in Paper I. Here, in the case of TW Hya, the \vsini\ is only about 6
\kms. Following the calculations in Paper II, an effective magnetic resolution of $\sim$ 1.5 kG is estimated. Motivated
by our previous results, we apply multi-component magnetic field models to TW Hya. We start with a two field 
component model to fit the data.

     The two field component model divides the surface of the star into three regions: two covered with magnetic fields 
of different strengths and filling factors, and the last region is field-free. The sum of the filling factors of the 
three components must be 1.0, and the spectra from the different regions are multiplied by their filling factors 
and summed to generate the final model spectrum. We again use the Marquardt technique
to find  magnetic field strengths and filling factors that produce the best fit to the observed line 
profiles. The two field component model yields a somewhat smoother fit to the observed line profiles with a smaller 
$\chi^2_r$ value of $1.29$. The resulting magnetic field strengths and filling factors for the best fit are 
$B_{1} = 3.0$ kG, $f_{1} = 0.56$ and $B_{2} = 6.5$ kG, $f_{2} = 0.16$. The mean field on the stellar surface for this
model is $\bar B = B_{1}f_{1} + B_{2}f_{2} = 2.72$ kG.  To estimate uncertainties caused by potential errors in the
atmospheric parameters derived from optical analysis, we re-analyze the IR spectra using a grid of assumed atmospheric
parameters in which we vary \Teff\ $\pm$ 400 K from the value in the default model, vary \logg\ down by 0.5 from
the default model, and vary [M/H] down by 0.5 from the default model. This results is a total of 12 different fits to
the observed line profiles.  Taking the mean and standard deviation of the resulting $\bar B$ values gives a new
estimate of $\bar B = 2.44 \pm 0.25$ kG on the stellar surface.

     In order to explore whether additional components improve the fit significantly, we repeated the analysis described
above for models with three or more components, each with a distinct field strength. For each model, we determined the
filling factors for each component by fitting the observed \ion{Ti}{1} line profiles, using the Marquardt algorithm.
Table \ref{multi} summarizes our results for three different multi-component models: M1 includes components with field
strengths of 0 -- 5 kG in steps of 1 kG, M2 includes components with field strengths of 1 -- 5 kG in steps of 2 kG, and
M3 includes components with field strengths of 0 -- 6 kG on steps of 2 kG. The last two rows of Table \ref{multi} give
the reduced $\chi^2$ and mean magnetic field over the surface for each model.  All three models produce good fits to the
data and indicate average field strength over the entire surface of 2.6 -- 2.8 kG.  Model M1 with 1 kG steps is
underconstrained, given the Zeeman resolution of 1.5 kG discussed above, leading to an anomalously low filling factor
for the 1 kG component, relative to the 0 and 2 kG components. Models M2 and M3 with 2 kG steps are well constrained.
Given our Zeeman resolution, we cannot tell whether the component with the weakest field has a characteristic field
strength of 0 kG or 1 kG.

\begin{figure}[ht]
  \begin{center}
  \includegraphics[scale=0.60]{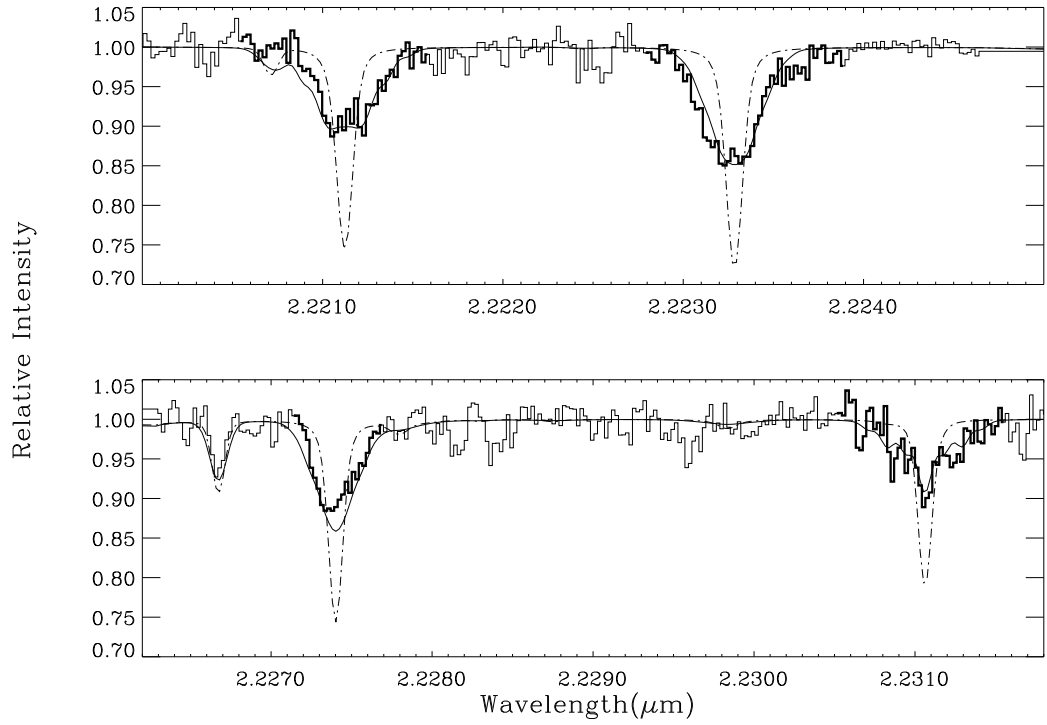}

   \caption{Best fit of infrared spectra using the multi-component
 magnetic field model M3. Magnetically sensitive Ti I lines are
fitted with the regions actually used in the fit shown again in bold
(histogram: data; dash-dotted line: fit without magnetic
field; smooth solid line: fit with magnetic broadening). }
   \label{irfig2}
   \end{center}
\end{figure}

Fig \ref{irfig2} compares model M3 with the observations.  The reduced $\chi^2$ for these models is similar to but slightly
larger than that found from the two field component model, suggesting that 2 components are adequate to fit the data
for TW Hya.  
The differing values of $\bar B$ derived from
all the models considered reflect the systematic uncertainty associated with the choice of the model used to analyze
the data.  Since models M1, M2, and M3 and the two field component model have similar $\chi_r^2$ values, we take the
mean and standard deviation of these as our best estimate for the average field on TW Hya: $\bar B = 2.71 \pm 0.08$
kG.  Above, we found that taking into account potential errors in the assumed stellar parameters resulted in an
additional 10\% uncertainty which we add in quadrature to that just given, to arrive at a final mean field estimate
of $\bar B = 2.71 \pm 0.28$.  Given that we have spent some considerable effort to derive accurate stellar
parameters for TW Hya, we believe this is a conservative estimate of the unceratinty in the mean magnetic field.

\subsection{Test of Analysis Technique}

As described above, we determined the magnetic properties of TW Hya by fitting the IR spectra of \ion{Ti}{1} lines in
the K band.  In this section, we use a Monte Carlo analysis to assess the stability of our analysis technique and the
possible impact of systematic errors.  We synthesize model spectra with a range of known magnetic properties, adding
noise comparable to the uncertainty in our actual observations, and then we analyze each synthetic spectrum using our
analysis technique.  To create simulated observations, we adopted stellar parameters appropriate for a typical K7
TTS similar to TW Hya, namely $\Teff = 4000$ K, $\logg = 3.5$, [M/H] = 0.0, and \vsini$ = 6.0$ \kms.

We analyze each simulated observation 100 times with different realizations of the noise and random initial guesses
for the magnetic field and its filling factor.  In all cases we analyze the simulated data assuming the same nonmagnetic
stellar parameters used to generate the artifical data.  Results are shown in Table \ref{table1}.  The first two columns
of the table give the input magnetic field strength and filling factor used to generate the simulated obsevation. The
third column gives the number of good runs out of 100 trials.  A bad runs occurs when the program returns $B$ values
trapped in a local minimum between 10 -- 20 kG due to noise in the data.  Their $\chi_r^2$ values are approximately twice
those of the ``good" fits, and plots of the fit show that they do not match the data.  The fourth and fifth columns
give average values of $B$ and $f$ for all good runs. The sixth and seventh columns give the ratio of $\bar B$ to its
input value and the corresponding standard deviation of this ratio.  Errors in measured $\bar B$ due to noise in the
data are typically 5\% or less, if \Teff, \logg, and [M/H] are accurate.

Padgett (1996) found that in a sample of 30 TTSs the effective temperatures determined from spectral classification
differ from the value obtained from measurements of line ratios in high resolution spectra.  The absolute value of the
difference averages 150 K with a scatter up to 600 K.  Additionally, gravities for TTSs usually come from placing stars
in the H-R diagram using uncertain distances and then comparing with uncertain pre-main sequence evolutionary tracks.
Although we determine \Teff\ and \logg\ spectroscopically, our atmospheric parameters could still be subject to error.
Thus, we use Monte Carlo tests to explore the effect of such errors on our derived magnetic parameters.  Table
\ref{table2} shows the results for a 200 K error in \Teff, while Table \ref{table3} shows the results for a 0.5 dex
error in \logg.  The Monte Carlo tests show that in the absence of accurate effective temperatures or gravities, the
resulting systematic errors in the derived mean field is about 10\% for typical field values and reasonable errors in
temperature or gravity.

\section{Discussion}
 
Using high resolution optical spectra, we determine the key atmospheric parameters for TW Hya which are 
listed in  Table \ref{optictab}. These values are generally in good agreement with previously published estimates.
The calibration between spectral type and \Teff\ given by Cohen and Kuhi (1979) is often used in studies of TTSs.  
This calibration is essentially that of Johnson (1966), which is based on IR photometric studies of cool main
sequence stars.  For K7 spectral type,  \Teff$ = 4000$ K with a quoted $2\%$ uncertainty.  Bessel (1991) also derives 
\Teff$ = 4000$ K for K7 stars by comparing broadband fluxes with blackbody distributions.  We obtain an effective 
temperature of $4126 \pm 24$ K for TW Hya.  From Papers I and II, BP Tau and Hubble 4 -- both K7 stars -- were found to 
have $\Teff = 4055 \pm 112$ K  and $\Teff = 4158 \pm 56$ K, respectively.  A statistical study of previous results for 
\Teff\ and spectral type for 268 stars by de Jager and Nieuwenhuijzen (1987) yields $\Teff = 4150 \pm 200$ K for K7
spectral type.

Previous results for the projected rotation velocity include $\vsini = 4 \pm 3$ \kms\ from Torres et al. (2000), 
$\vsini < 6.0$ \kms\ from Johns-Krull and Valenti (2001), and $5 \pm 2$ \kms\ from Alencar and Batalha (2002). All agree 
well with our value of $5.8 \pm 0.6$ \kms.  To our knowledge, there has been no previous determination of the metallicity
of TW Hya.  Our value of [M/H] $= -0.11 \pm 0.12 $ is well within the range of [Fe/H] values reported by Padgett (1996)
in her study of several TTSs.  For \logg, Alencar and Batalha (2002) obtain a value of $4.44 \pm 0.05$ using the line 
synthesis models of Barbuy (1982) and Valenti and Piskunov (1996).  Our value of $\logg = 4.84 \pm 0.16$ is somewhat high,
and differs from this estimate by $2.4\sigma$.  The sensitivity to gravity in our analysis comes from the depth of the
TiO bandhead (which is also quite temperature sensitive) as well as from the
two strong lines at 6493.8 \AA\ and 6495.0 \AA\ which show damping wings.  We generally avoid more sensitive
gravity diagnostics such as the \ion{Na}{1} D lines since their profiles in CTTSs are usually contaminated by emission 
and/or absorption from circumstellar material.  In the previous section, we show that this difference of 0.40 in \logg\
is expected to produce no more than a 10\% error in our mean magnetic field estimate, which we account for in our
quoted uncertainties.

TW Hya has a luminosity of $\log(L_* / L_\odot) = -0.62 \pm 0.11$, where the uncertainty in luminosty is dominated
by uncertainty in the Hipparcos parallax.  Combining this luminosity with our spectroscopically determined \Teff\
yields a stellar radius of $R_* = 0.96 \pm 0.12 {\rm R}_\odot$.  This radius together with our spectroscopic
gravity estimate yields a mass of $M_* = 2.31 \pm 0.69 {\rm M}_\odot$, where the uncertainty in our spectroscopic
mass has roughly equal contrbutions from the uncertainties in \logg\ and $R_*$.  Comparing the location of TW Hya
on an HR Diagram with pre-main sequence evolutionary tracks of D'Antona \& Mazzitelli (1994) and Palla \& Stalher
(1999) yields stellar masses of $M_* = 0.79 {\rm M}_\odot$ and $0.80 {\rm M}_\odot$ respectively.  Our spectroscopic
mass is substantially larger than that estimated from evolutionary tracks and reflects our high value of \logg.
Given the large uncertainty in our spectroscopic mass, the discrepancy is only $2.2\sigma$.  If we instead use the
\logg\ reported by Alencar and Batalha (2002), we find $M_* = 0.93 \pm 0.10$, in better agreement with evolutionary
tracks.

Papers I and II determined spectroscopic and evolutionary track masses for BP Tau and Hubble 4.  While the difference in
any individual case is not statistically significant, it is interesting that in all three analyses to date, the
spectroscopic masses are larger than masses based on evolutionary tracks.
For stars with mass less than or equal to 1.2M$_\odot$, Hillenbrand and White (2004) found a similar trend in their
study of stars with dynamically determined masses.  The same result is also found by Reiners, Basri, and Mohanty (2005)
in their analysis of a pre-main sequence double lined spectroscopic binary in Upper Scorpious.  Therefore, it is likely
that current pre-main sequence evolutionary tracks do indeed systematically underestimate the mass of young stars.
For main-sequence stars, Valenti and Fischer (2005) find an analogous discrepancy between spectroscopic and isochrone
masses, though the magnitude of the discrepancy is typically only 0.1 dex.

On the other hand, the mass we derive based on our gravity estimate is quite high relative to these evolutionary
tracks.  In order to explore the effect of this on our analysis, we use $R_*$ determined above and $M_*$ from evolutionary 
tracks to calculate \logg$ = 4.38$.  With this \logg, we reanalyze our optical spectra and obtain
another set of atmospheric parameters: $\Teff = 4110$ K, [M/H] $= -0.23$, $\vsini = 6.2$ \kms, and \textit{r} = $0.61$.
This fit returns a slightly larger $\chi^2_r$ and is shown as the dash-dotted line in Figure \ref{opticfig}.
In addition, we fit the single magnetic component model with these atmospheric parameters and find $\bar B = 2.07$ kG.
This fit returns $\chi^2_r = 1.88$ and should be compared with our best fit single field component model with our 
previously determined atmospheric parameters which yields $\bar B = 2.21$ kG and $\chi^2_r = 1.70$.  The new atmospheric
parameters produce a $\sim 6$\% difference in the resulting mean field, confirming the results found above which showed
that reasonable errors in the assumed gravity produce $\sim 10$\% or less error in the derived magnetic field 
properites. 

In Paper I, Johns--Krull et al. examine the magnetospheric accretion models of K\"onigl (1991), 
Cameron \& Campbell (1993), and Shu et al. (1994) and summarize analytic equations for the predicted surface 
magnetic field strength of each model as a function of stellar mass, radius, rotation rate, and disk
accretion rate.  Using equations (2) -- (4) of Paper I, and adopting an accretion rate of $\dot M = 2 \times 10^{-9} 
{\rm M}_\odot$ yr$^{-1}$ from Herczeg et al. (2004), a rotation period $P = 2.2$ days from Mekkaden (1998), 
along with the stellar radius (0.96 R$_\odot$) and mass (2.31 M$_\odot$) estimated above, we estimate $B_*$. 
The K\"onigl (1991) model gives $B_* = 2.78$ kG, while the Cameron \& Campbell (1993) model gives $B_* = 0.60$ kG,
and the model of Shu et al. (1994) predicts $B_* = 2.74 $ kG.  Our measured $\bar B = 2.71 \pm 0.28$ kG is very close
to the predictions K\"onigl (1991) and Shu et al. (1994), while it is larger than the prediction of Cameron and Campbell
(1993).  However, we caution that our measurement is just the unsigned field {\it strength};
we are unable to comment on the magnetic field geometry, which theories currently assume is dipolar.

Combining our measured average field strength with our calculated stellar radius, we can estimate the total stellar 
magnetic flux on the star.  Pevtsov et al. (2003) find a nearly linear relationship between total unsigned magnetic flux  
and X-ray luminosity, $L_X$, for solar features (ranging from the smallest scales up to the entire Sun) and dwarf stars. 
Using this relationship, we predict $L_X = 1.41 \pm 0.14 \times 10^{30}$ erg s$^{-1}$.  This value agrees well with the 
observed values of $0.90 - 1.76 \times 10^{30}$ erg s$^{-1}$ from  Kastner et al. (2002) assuming a distance of 56 pc. 

We can also use our magnetic field measurements to test the concept of pressure equipartition in the stellar photosphere.
Spruit \& Zweibel (1979) computed flux tube equilibrium models, showing that magnetic field
strength is expected to scale with gas pressure in the surrounding
non-magnetic photosphere.  Field strengths set by pressure equipartition 
appear to be observed in G and K dwarfs (e.g. Saar 1990, 1994, 1996) and 
possibly in M dwarfs (e.g. Johns--Krull \& Valenti 1996).  Most TTSs have 
relatively low surface gravities and hence low photospheric gas pressures,
so that equipartition flux tubes would have relatively low 
magnetic field strengths compared to cool dwarfs.
Indeed, Safier (1999) examined in some detail the ability of TTS photospheres
to confine magnetic flux tubes via pressure balance with the surrounding 
quiet photosphere, concluding that the maximum field strength allowable on
TTSs is substantially below the current detections.  On the other hand, it is doubtful
that pressure equipartition arguments should hold when the magnetic field
filling factor approaches unity on the stellar surface.  Saar (1996)
finds that the magnetic field strength does rise above the pressure
equipartition value for very active K and M dwarfs once the filling factor
reaches a value near one.  Papers I and II found that the mean fields on BP Tau and
Hubble 4 are too large to be confined by gas pressure in the surrounding photosphere,
suggesting that magnetic fields cover the entire surface.  

TW Hya is an interesting case because its older age and corresponding
higher surface gravity imply higher equipartition magnetic field strengths.  Using the
atmospheric parameters from Table 2, we find that the photospheric gas pressure in
the atmosphere at the level where the local temperature is equal to the effective
temperature implies a maximum field strength of 2.78 kG.  Adopting the lower
gravity of $\logg = 4.38$ and the associated temperature and metallicity described
above gives a field strength of 2.14 kG at this same level in the stellar atmosphere.
The first estimate is very similar to the derived mean field while the second is
only somewhat lower.  The two field component model and the multi-component models
M1 - 3 all indicate that substantial portions of the stellar surface is covered by
fields larger than these equipartition estimates; however, it is not certain that
these observations demand unit filling factor of field regions in the case of TW Hya.  In the analysis
above, it was assumed that the atmospheric structure of the magnetic regions is
identical to that in the non-magnetic regions (the magnetic regions are neither
hot plage-like regions nor cool spot-like regions).  In the Sun, the field strength in
cool spots is a factor of $\sim 2$ larger than the typical value found outside of 
spots.  As discussed in Paper II (see also Guenther et al. 1999), cool spot-like 
regions (and to a lesser extent hot plage-like regions) can affect the derived filling 
factor of these fields, but have a
realtively weak effect on the derived field strength when actual Zeeman broadening
is detected.  The strong field regions on TW Hya are only about a factor of 2
stronger than the pressure equipartition field value, and the majority of the
field found on TW Hya is near this equipartition value.  Therefore, on TW Hya, the
strong fields may be confined to cool spots while the weaker fields exist in 
plage-like regions on the surface of the star, and there is no need to assume unit filling
factor of magnetic field regions.  
 
Since TTSs generally are heavily spotted stars (e.g. Herbst et al. 1994 and references therein; 
Strassmeier, Welty, \& Rice 1994 and references therein), there may be a need to model the observed
spectra with 2 atmospheric components, each with a
different effective temperature.  In sunspots (and presumably starspots), magnetic pressure partially evacuates the
spot, changing its density structure relative to the surrounding atmosphere.  Unfortunately, there are no models for 
the atmospheric structure
of spots on stars other than the Sun, and constructing such models is well beyond the scope of this paper.  One
could imagine trying to use model atmospheres simply scaled down in effective temperature (keeping the gravity
and metallicity fixed) to begin to explore 
magnetic fields in spots and the resulting effect they might have on magnetic field analysis on TTSs.  Paper II
did just this for Hubble 4, finding that models with spots which were constrained to be 1000 K cooler than
the non-spotted atmosphere had more difficulty simultaneously matching the optical and IR spectra than did models 
without a spot contribution.  The analysis of Hubble 4 also showed that the derived mean field from the spot model
was within 18\% of the derived mean field from the single component fit, with the change entirely due to the
derived filling factor.  While spots contribute relatively more to the continuum in the IR than in the optical,
the strong temperature sensitivity of the TiO bandhead at 7055 \AA\ results in significant contributions to this
feature from spots and is a key constraint for any model.
Building on these results, we restricted the modeling in this paper to a single effective temperature model.  We
note though that our predicted CO line strengths are fairly close to the observed line strengths, and that lowering
the effective temperature by 300 K actually produces a slightly higher $\chi^2_r$ for the CO line profiles.  
Evidently, the K band spectrum on TW Hya is not dominated by a substantially cooler (spot) component on the star.

Realistic models of spot atmospheres are needed to make further progress on
testing just how much cool spots may contribute to the magnetic fields we have
detected on TW Hya and other TTSs. In Paper II a spot model was used, but it 
was simply the atmosphere of a cooler star, with the same gravity and metallicity
as that found for Hubble 4.  On the Sun, magnetic pressure support
partially evacuates the gas in a spot, changing its density compared
to the quiet atmosphere.  Such effects need to be considered for
TTS (and other cool stars such as dMe stars).  In addition, since it
appears that strong magnetic fields on TTSs are also present in a substantial
portion of the non-spotted photosphere, atmospheric models taking into account the
magnetic pressure support should also be considered for this component of
the atmosphere.  Such a modeling effort is well beyond the scope of
the current paper, but these results should be regarded as an appeal to 
theoreticians to begin to consider such problems, particularly since the 
data presented here clearly indicates the presence of strong magnetic fields
on a large portion of TW Hya.  Such strong fields have also been inferred
for a handful of other TTSs (Basri et al. 1992, Guenther et al. 1999, 
Papers I and II), and these strong fields are likely to be a common feature of TTSs
given their strong X-ray emission (e.g. Feigelson et al. 2003).
Incorporating the effects of the magnetic field on the atmospheric
structure will likely have important consequences beyond simply
determing more accurate magnetic field parameters on TTSs.  Perhaps
it is the influence of these fields on the atmospheric structure that
is responsible for the inconsistent mass determinations for TTSs
described above?  Future work is needed to know for sure.

\section{Conclusion}

We analyzed high resolution optical spectra of the classical T Tauri star TW Hya. We successfully fitted
numerous atomic and molecular line features in 3 wavelength regions,
which are modeled well in both the solar spectrum and the spectrum of 61 Cyg B, a K7V reference star. 
In the process we determined the  key atmospheric parameters (\Teff, \logg, [M/H], \vsini, and veiling) of the star.

With these atmospheric parameters, we simultaneously fit 4 magnetically sensitive \ion{Ti}{1} lines at 2.2 $\mu$m 
to look for Zeeman broadening. We also fit several magnetically insensitive CO lines at 2.3 $\mu$m to 
verify that the parameters determined from optical analysis accurately reproduce the IR spectrum of TW Hya. 
The CO model spectrum agrees well with the data, and the fit to \ion{Ti}{1} lines using various magnetic
models yields a surface magnetic field strength of $\bar B = 2.71\pm0.28$ kG, averaged over the entire
surface of TW Hya.

We tested the sensitivity and stablity of our analysis to 
systematic errors in the effective temperature and gravity, 
finding that errors of 200 K in $\Teff$   
or 0.5 dex in $\logg$ only result in a systematic error of $\sim 10\%$ in the derived mean magnetic field. 
Our measured field strength on TW Hya agrees well with the field strength expected from magnetospheric accretion models.
However, we caution that these models assume a dipolar field geometry, and the data presented here do not 
constrain the field geometry on TW Hya. Circular polarization measurements, on the other hand, would provide
strong constraints on the dipole field geometry of TW Hya.

\acknowledgements
CMJ-K and HY would like to acknowledge partial support from the NASA Origins of Solar Systems 
program through grant number NAG5-13103 made to Rice University.

\clearpage



\clearpage


\begin{table}
  \caption {Journal of observations.}\label{obs}
  \begin{center}
  \leavevmode
  \footnotesize
    \begin{tabular}[h]{cccccc}
    \tableline\tableline
   \ Wavelength Setting & Main Lines & UT date & UT time & Observatory & Total
    exposure time(s)  \\[+5pt]
    \tableline

     2.2228  $\mu$m & \ion{Ti}{1} & 16-Apr-98  & 06:47 & IRTF & 3200 \\[+5pt]
     2.2300  $\mu$m & \ion{Ti}{1} & 14-Apr-98  & 07:39 & IRTF & 4500 \\[+5pt]
     2.31294 $\mu$m &    CO       & 15-Apr-98  & 07:37 & IRTF & 6300 \\[+5pt]
   5803 -- 7376 \AA &   TiO       & 05-Apr-99  & 07:28 & McDonald &
10800 \\[+5pt]
   5803 -- 7376 \AA &   TiO       & 06-Apr-99  & 07:10 & McDonald &
7200 \\[+5pt]

    \tableline
  \end{tabular}
 \end{center}
\end{table}

\begin{table}
  \caption{ Best fit atmospheric parameters for optical spectra.}
  \label{optictab}
  \begin{center}
  \leavevmode
   \footnotesize
    \begin{tabular}[h]{cccccc}
    \tableline\tableline
                & \Teff & \logg         &       & \vsini &          \\[+5pt]
          Param & (K)   & (cm s${-2}$)  & [M/H] & (\kms) & Veiling  \\[+5pt]
    \hline \\[-5pt]
          Value &  4126 &          4.84 & -0.11 &  5.80  &     0.61 \\[+5pt]
       $\sigma$ &    24 &          0.16 &  0.13 &  0.63  &     0.15 \\[+5pt]
    \hline \\
    \end{tabular}
 \end{center}
\end{table}

\begin{table}
  \caption{Multi-component magnetic fits.} 
   \label{multi}
  \begin{center}
   \leavevmode
   \begin{tabular}[h]{lccc}
   \tableline\tableline

  \                   &   M1    &    M2   & M3          \\ \tableline
$f_{0{\rm kG}}$       & 0.20    & \nodata & 0.18        \\
$f_{1{\rm kG}}$       & 0.04    & 0.32    & \nodata     \\
$f_{2{\rm kG}}$       & 0.20    & \nodata & 0.38        \\
$f_{3{\rm kG}}$       & 0.28    & 0.45    & \nodata     \\
$f_{4{\rm kG}}$       & 0.12    & \nodata & 0.32        \\
$f_{5{\rm kG}}$       & 0.16    & 0.23    & \nodata     \\
$f_{6{\rm kG}}$       & \nodata & \nodata & 0.11        \\
$\chi^2_r$            & 1.32    & 1.34    & 1.31        \\
$\bar B$ (kG)         & 2.6     & 2.8     & 2.7         \\ \tableline
\end{tabular}
\end{center}
\end{table}

\begin{table}
  \caption{Test results for fitting artificial data created assuming \Teff\ =4000 K, \logg\ = 3.5,
  [M/H] = 0, and $\vsini = 6.0$ \kms. Artificial data is fit assuming the same atmospheric parameters
  as those used to create the data.}
  \label{table1}
  \begin{center}
    \leavevmode
    \footnotesize
    \begin{tabular}[h]{ccrrccc}
      \tableline\tableline
    \multicolumn{2}{c}{Model}&  good & \multicolumn{3}{c}{Aver. of Results} & \\[+5pt]
      $B$($G$) & $f$ & runs &  $B$($G$)    &  $f$  &
      $Bf$/$Bf_{\rm exp}$ & $\sigma $   \\[+5pt]
      \hline \\[-5pt]
    1000  &  0.25   &    88     &    911.14  &  0.23   &  0.802      & 0.0549     \\
    1000  &  0.50   &    98     &    1001.6  &  0.52   &  1.032      & 0.0028     \\
    1000  &  0.75   &    97     &    931.08  &  0.80   &  0.988      & 0.0011     \\
    2000  &  0.25   &   100     &    1823.9  &  0.30   &  1.079      & 0.0051     \\
    2000  &  0.50   &   100     &    1866.3  &  0.52   &  0.977      & 0.0062     \\
    2000  &  0.75   &   100     &    2092.9  &  0.72   &  1.005      & 0.0004     \\
    3000  &  0.25   &   100     &    3005.3  &  0.23   &  0.917      & 0.0103     \\
    3000  &  0.50   &   100     &    3020.6  &  0.50   &  1.017      & 0.0008     \\
    3000  &  0.75   &   100     &    2975.7  &  0.75   &  0.996      & 0.0003     \\
    4000  &  0.25   &    98     &    4581.8  &  0.25   &  1.089      & 0.1532     \\
    4000  &  0.50   &   100     &    4185.4  &  0.51   &  1.051      & 0.0759     \\
    4000  &  0.75   &    99     &    4173.2  &  0.74   &  1.005      & 0.0920     \\
      \hline \\
      \end{tabular}
  \end{center}
\end{table}

\begin{table}[bht]
  \caption{Test results for fitting artifical data created assuming \Teff\ = 4000 K, \logg\ = 3.5,
  [M/H] = 0, and $\vsini = 6.0$ \kms. Artificial data is fit assuming \Teff\ = 4200 K, \logg\ = 3.5,
  [M/H] = 0, and $\vsini = 6.0$ \kms.}
  \label{table2}
  \begin{center}
    \leavevmode
    \footnotesize
    \begin{tabular}[h]{ccrrccc}
      \tableline\tableline
    \multicolumn{2}{c}{Model}&  good & \multicolumn{3}{c}{Aver. of Results} & \\[+5pt]
      $B$($G$) & $f$ & runs &  $B$($G$)    &  $f$  &
      $Bf$/$Bf_{\rm exp}$ & $\sigma $   \\[+5pt]
      \hline \\[-5pt]
 1000   & 0.25     &  98   &  493.81  &  0.97    & 1.840 &    0.1172     \\
 1000   & 0.50     &  97   &  747.11  &  0.98    & 1.461 &    0.0321     \\
 1000   & 0.75     & 100   &  962.16  &  0.96    & 1.185 &    0.1310     \\
 2000   & 0.25     & 100   &  1074.1  &  0.61    & 1.303 &    0.0047     \\
 2000   & 0.50     &  99   &  1568.4  &  0.75    & 1.181 &    0.0023     \\
 2000   & 0.75     & 100   &  1737.5  &  0.96    & 1.112 &    0.0255     \\
 3000   & 0.25     & 100   &  2335.1  &  0.32    & 0.982 &    0.0029     \\
 3000   & 0.50     & 100   &  2773.1  &  0.60    & 1.111 &    0.0005     \\
 3000   & 0.75     & 100   &  2790.3  &  0.92    & 1.138 &    0.0128     \\
 4000   & 0.25     & 100   &  3803.4  &  0.26    & 0.986 &    0.0006     \\
 4000   & 0.50     &  99   &  3818.8  &  0.56    & 1.073 &    0.0617     \\
 4000   & 0.75     & 100   &  3866.5  &  0.90    & 1.157 &    0.0070     \\
      \hline \\
      \end{tabular}
  \end{center}
\end{table}

\begin{table}[bht]
  \caption{Test results for fitting artifical data created assuming \Teff\ = 4200 K, \logg\ = 3.5,
  [M/H] = 0, and $\vsini = 6.0$ \kms. Artificial data is fit assuming \Teff\ = 4200 K, \logg\ = 4.0,
  [M/H] = 0, and $\vsini = 6.0$ \kms.}
  \label{table3}
  \begin{center}
    \leavevmode
    \footnotesize
    \begin{tabular}[h]{ccrrccc}
      \tableline\tableline
    \multicolumn{2}{c}{Model}&  good & \multicolumn{3}{c}{Aver. of Results} & \\[+5pt]
      $B$($G$) & $f$ & runs &  $B$($G$)    &  $f$  &
      $Bf$/$Bf_{\rm exp}$ & $\sigma $   \\[+5pt]
      \hline \\[-5pt]
 1000   & 0.25     &   89 &      7819.6 &     0.05  &  1.579 &   0.2942 \\
 1000   & 0.50     &   86 &      1660.9 &     0.19  &  0.629 &   0.0631 \\
 1000   & 0.75     &  100 &      1547.2 &     0.31  &  0.640 &   0.0380 \\
 2000   & 0.25     &  100 &      3580.8 &     0.16  &  1.011 &   0.1441 \\
 2000   & 0.50     &  100 &      2417.3 &     0.38  &  0.920 &   0.0004 \\
 2000   & 0.75     &  100 &      2180.9 &     0.61  &  0.888 &   0.0385 \\
 3000   & 0.25     &   98 &      4386.1 &     0.23  &  1.236 &   0.5194 \\
 3000   & 0.50     &  100 &      3279.8 &     0.48  &  1.030 &   0.1015 \\
 3000   & 0.75     &  100 &      3208.0 &     0.67  &  0.925 &   0.0554 \\
 4000   & 0.25     &   99 &      4635.2 &     0.24  &  1.113 &   0.2760 \\
 4000   & 0.50     &   99 &      4221.4 &     0.47  &  0.972 &   0.1333 \\
 4000   & 0.75     &  100 &      4164.3 &     0.69  &  0.930 &   0.1459 \\
      \hline \\
      \end{tabular}
  \end{center}
\end{table}

 \clearpage

\clearpage

\clearpage

\clearpage

\end{document}